\documentclass[sigconf]{acmart}

\AtBeginDocument{%
  }

\usepackage{listings}
\usepackage{lstcustom}

\usepackage{todonotes}
\usepackage{enumitem}
\usepackage[normalem]{ulem}
\usepackage{booktabs}  
\usepackage{balance}
\usepackage{listings, xcolor}
\usepackage{multicol}
\usepackage{multirow}
\usepackage{multitoc}
\usepackage{textcomp}
\usepackage[ruled,vlined]{algorithm2e}
\usepackage{graphicx}
\usepackage{hyperref}
\usepackage{float}
\usepackage{algorithmic}
\usepackage{soul}
\usepackage{caption}
\usepackage{subcaption}
\usepackage{tikz}
\usepackage{pifont}
\usepackage{microtype}
\usepackage{siunitx}
\usepackage{gensymb}
\usepackage{array}

\begin{document}

\title{Accelerating Java Ray Tracing Applications on Heterogeneous Hardware}

\author{Vinh Pham Van}
\affiliation{%
  \institution{Improbable Worlds Ltd.}
  \country{United Kingdom}
  \postcode{M13 9PL}
}

\author{Juan Fumero}
\affiliation{%
  \institution{The University of Manchester}
  \city{Manchester}
  \country{United Kingdom}
  \postcode{M13 9PL}
}
\email{juan.fumero@manchester.ac.uk}

\author{Athanasios Stratikopoulos}
\affiliation{%
  \institution{The University of Manchester}
  \city{Manchester}
  \country{United Kingdom}
  \postcode{M13 9PL}
}
\email{{first}.{last}@manchester.ac.uk}

\author{Florin Blanaru}
\affiliation{%
  \institution{The University of Manchester}
  \city{Manchester}
  \country{United Kingdom}
  \postcode{M13 9PL}
}
\email{florin.blanaru@octoml.ai}

\author{Christos Kotselidis}
\affiliation{%
  \institution{The University of Manchester}
  \city{Manchester}
  \country{United Kingdom}
  \postcode{M13 9PL}
}
\email{christos.kotselidis@manchester.ac.uk}

\renewcommand{\shortauthors}{V. Pham Van, J. Fumero, A. Stratikopoulos, F. Blanaru, C. Kotselidis}

\begin{abstract}
Ray tracing has been typically known as a graphics rendering method capable of producing highly realistic imagery and visual effects generated by computers.
More recently the performance improvements in Graphics Processing Units (GPUs) have enabled developers to exploit sufficient computing power to build a fair amount of 
ray tracing applications with the ability to run in real-time.
Typically, real-time ray tracing is achieved by utilizing high performance kernels written in CUDA, OpenCL, and Vulkan which can be invoked by high-level languages via native bindings; a technique that fragments application code bases as well as limits portability.
    
This paper presents a hardware-accelerated ray tracing rendering engine, fully written in Java, that can seamlessly harness the performance of underlying GPUs via the TornadoVM framework.
Through this paper, we show the potential of Java and acceleration frameworks to process in real time a compute intensive application.
Our results indicate that it is possible to enable real time ray tracing from Java by achieving up to 234, 152, 45 frames-per-second in 720p, 1080p, and 4K resolutions, respectively.
\end{abstract}

\ccsdesc[300]{General and reference~Computing standards, RFCs and guidelines}
\ccsdesc[500]{Applied computing~Publishing}

\ccsdesc[500]{Software and its engineering~Software libraries and repositories}

\keywords{FPGAs, High-Performance, Java, JIT compilation, Optimizations}


\maketitle

\section{Introduction}

\begin{figure}[b!]
\centering
  \includegraphics[width=\linewidth]{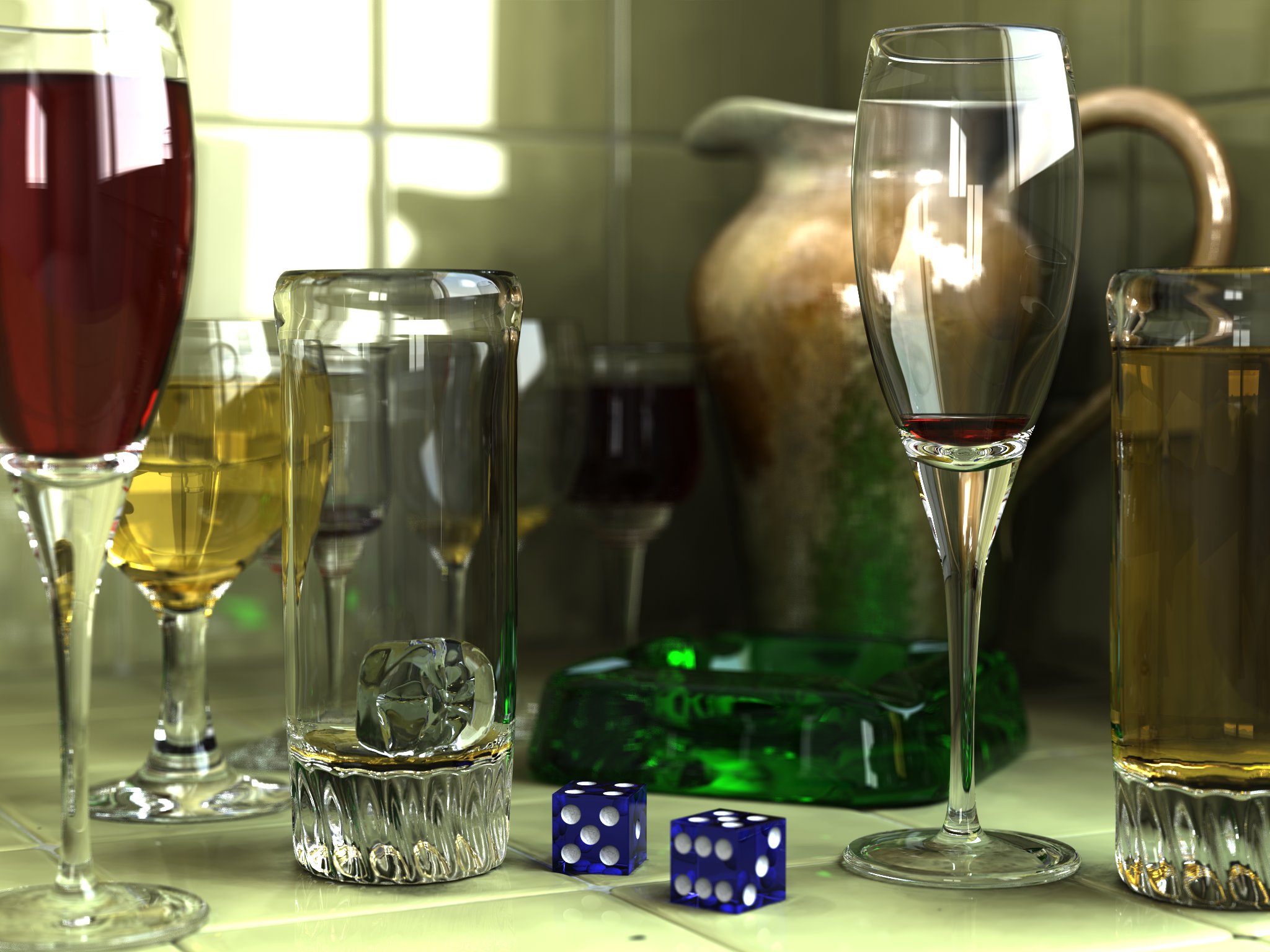}
  \caption{Photo-realistic scene rendered through ray tracing~\cite{glasses:2006}.}
  \label{fig:glasses}
\end{figure}

Ray tracing~\cite{10.1145/1468075.1468082} is a rendering technique that has gained significant traction within the field of computer graphics for its ability to create believable, photo-realistic computer-generated imagery. 
By modelling the way light rays behave in the real world, environments can be simulated with accurate portrayals of shadows and reflections alongside a variety of 
optical effects that exist in nature, as shown in Figure~\ref{fig:glasses}.

Heavy computational demands of ray tracing for a long time have predominantly limited its usages to the film and television industries 
where high levels of processing power are available to produce pre-rendered scenes and visual effects~\cite{christensen:2016}.
However, recently, the increasing performance of graphics processing units (GPUs) have enabled access to publicly available hardware acceleration that provide
 sufficient processing power to enable ray tracing within real-time applications, such as video games, 
 where frames need to be consecutively rendered within a matter of milliseconds~\cite{deng:2017}.

Heterogeneous hardware accelerators, such as GPUs, have brought forth a heterogeneous programming landscape that poses several challenges compared to traditional software development. 
Their programmability 
require adoption of specialized low-level programming languages, such as OpenCL~\cite{opencl:2022} or CUDA~\cite{cuda:2022}, 
in which deep understanding of architectural details in relation to the parallel programming models  can result in increasing development expenses and time-to-market~\cite{bilavarn:hal-01287829}. 
Besides, the portability of accelerated, high-performance code is limited due to the re-factorizations needed whenever newer generations of devices are introduced, leading to additional maintenance costs~\cite{sathre:2019}.

High-level programming languages, such as Java and Python, were designed to abstract low-level concepts,
but have been generally avoided for heterogeneous programming due to their lack of native support for hardware acceleration.
State-of-the-art attempts at high-level, GPU-accelerated ray tracing, such as Mamba Tracer~\cite{mamba-tracer:2022} and Python RTX~\cite{python-rtx:2018}, only work through invocations of manually implemented OpenCL or CUDA code to perform heavy computations.
As alternatives, Aparapi~\cite{aparapi:2022}, Marawacc~\cite{10.1145/2807426.2807428, Fumero17:AIPGPUs}, and TornadoVM~\cite{tornadovm:2022} are parallel programming frameworks designed to exploit the computational power of heterogeneous hardware (e.g., GPUs) in a transparent manner. 

In this work, we explore a high-level approach to implement hardware accelerated ray tracing with the potential to massively outperform CPU-based implementations and run in real-time, while written entirely in Java.
Our open-source implementation\footnote{https://github.com/Vinhixus/TornadoVM-Ray-Tracer} does not contain any low-level programming or platform-specific optimizations, and it is fully implemented in Java using the TornadoVM APIs.
More specifically, we develop a ray tracing application from the ground up of the entire rendering pipeline, synthesizing a scene containing primitive shapes with displays of ray traced reflections and shadows. 
In detail, this paper makes the following contributions:

\begin{enumerate}
  \item It develops an interactive rendering engine that produces scenes with ray traced optical effects in real-time fully implemented in Java. 
  \item It analyzes how TornadoVM allows users to exploit the high-performance heterogeneous hardware at high-level, and accelerates the computations of ray tracing algorithms by transparently using commodity GPUs.
  \item It performs a performance evaluation of the proposed ray tracing engine across different types of accelerators and frame sizes, showcasing real-time rendering of 234, 152, 45 frames-per-second in 720p, 1080p, and 4K resolutions respectively. 
\end{enumerate}

The remaining of the paper is structured as follows:
Section~\ref{section:background} introduces the concept of ray tracing and hardware acceleration focusing on TornadoVM.
Section~\ref{section:design} describes the design of our framework and explains the operations used to process, in real time, shadows, textures and reflections. 
Section~\ref{section:evaluation} presents the performance evaluation and, finally, Sections~\ref{section:relatedwork} and ~\ref{section:conclusions} present the related work and conclusions, respectively.
\section{Background}
\label{section:background}

In computer graphics, \textit{rendering} is the process responsible for taking a three dimensional \textit{scene} containing various geometric objects and projecting them based on a viewing perspective onto a 2-dimensional representation to be displayed on monitor screens \cite{computer-graphics:2016}.
The main objective of rendering is to produce an accurate recreation of how these 3D objects would appear within the same environment in real life.

\subsection{Rendering techniques}
As follows, we describe the main components of a ray tracer application and common techniques. 

\subsubsection{The appearance of the real physical world}
\label{section:light-behavior}

\textit{Light} is probably the most important aspect of the physical appearance in the real world. Light consists of a set of light rays, a wave-like stream of photon particles that carry color information in wavelengths, that originate from sources such as a lamp or the sun. These light rays bounce around the environment, colliding with objects, where a number of interactions may occur depending on the shape and material of the object surface \cite{sliney:2016}:

\vspace{-0.2cm}\paragraph{Absorption}
The object may absorb the light, terminating its progress; 100\% absorption however, does not exist in nature, realistically only a percentage of the light is absorbed depending on the material.

\vspace{-0.2cm}\paragraph{Reflection}
The light may get reflected in a number of directions. A fully reflective surface, such as a mirror, reflects all of the light in one direction symmetrical to the incoming light ray, whereas a diffuse surface such as plastic could disperse the light in many directions with varying intensities.

\vspace{-0.2cm}\paragraph{Refraction}
In the case that the object is translucent, such as a glass or a body of water, a percentage of the incoming light ray may pass through the surface, allowing for the object to produce a see-through effect.

\vspace{-0.2cm}\paragraph{Fluorescence}
Some objects, such as the mineral rock ruby, may absorb the energy of the light ray and re-emit it with different properties, such as a different color.

Eventually, after a number of interactions, a light ray may end up at our eyes, where a set of photo-receptors detect the wavelength of the ray, which our brain processes as a color, producing an image which results in a perception of the world \cite{sliney:2016}.

\subsubsection{An overview of ray tracing}

\begin{figure}[t!]
\centering
  \includegraphics[width=\linewidth]{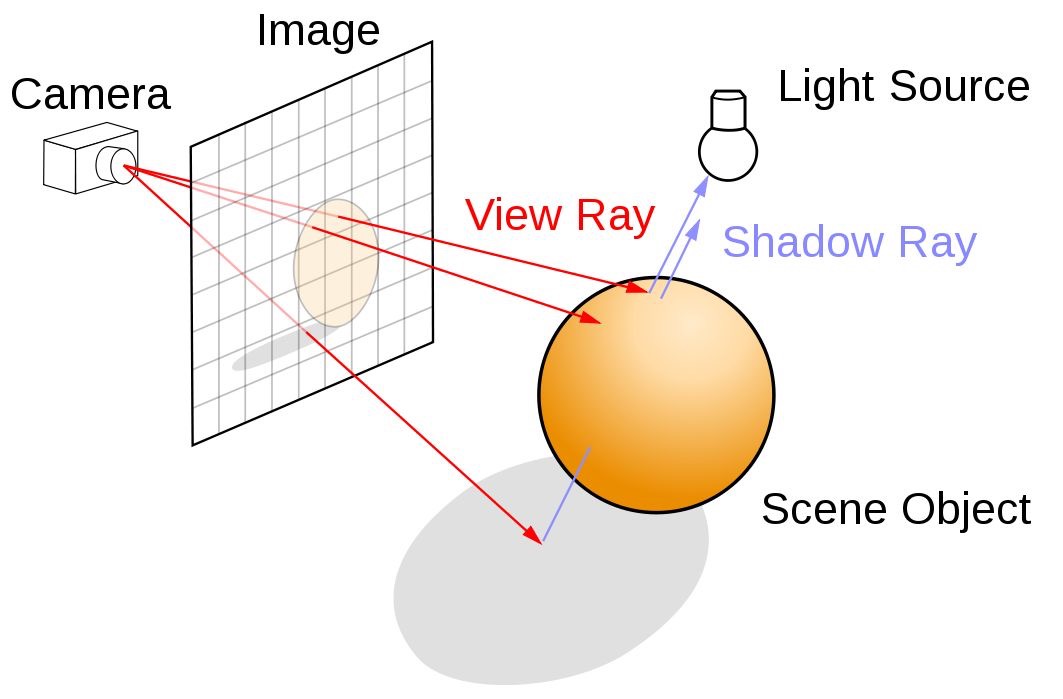}
  \caption{Arthur Appel's ray tracing basics.}
  \label{fig:ray-tracing}
  \vspace{-1.5em}
\end{figure}

While modelling the exact real world behavior of light would be the key to rendering photo-realistic images, it quickly becomes apparent that simulating and tracing every light ray from those starting off from the light sources to the ones generated at every interaction with objects produces an overwhelming amount of rays  do not reach an observer and consequently do not contribute to the final image.
To eliminate the redundant computations, researchers over the years have made efforts to identify the rays within a scene that contribute to the final picture and separate them from the ones that can be omitted.
The first idea that built the basis of ray tracing was proposed by Arthur Appel in 1968 \cite{appel:1968}.
Since the only light rays of interest are the ones reaching the eye of the observer, Appel suggested to generate rays starting from the view point and trace its path and behavior backwards, essentially reversing the entire real-life process. Appel used ray tracing to determine primary visibility, in other words to find the closest surface to the camera at each pixel, and trace secondary \textit{shadow rays} to the light source to define whether the point was in a cast shadow or not (Figure \ref{fig:ray-tracing}).

Not long after, in 1971 Robert A. Goldstein and Roger Nagel \cite{goldstein:1971} extended the works of Appel by not only producing cast shadows, but also computing the surface normal (the normal vector at a given surface point) at an intersection point to determine the direction the surface is facing. 
Knowing the position of the light source, it could then be established whether the given surface point faces towards or away from the light, according to which the brightness of the color could be adjusted to produce a shaded effect.
In 1980, Turner Whitted \cite{whitted:1980} proposed the modelling of light bounces from the interactions outlined in the previous section (\ref{section:light-behavior}). 
Whitted continued the works of Appel, Goldstein and Nagel by generating secondary \textit{reflection and refraction rays} at intersection points, which were traced throughout the scene to compute colors, producing the effects of reflections and translucent objects. 
This process is recursively continued until the light ray exits the scene or runs out of energy.

Turner Whitted's work became the baseline form of modern ray tracing techniques, known as the \textit{Classical Whitted-style Ray-Tracer} \cite{whitted:1980}, which modelled enough of the real-world behavior of light to produce extremely realistic visuals.

\subsubsection{The parallel nature of ray tracing}

One extremely advantageous property of ray tracing is that it is highly parallelizable. This is due to the fact that the entire process that obtains the color for one pixel is completely independent from another pixel.
These kinds of algorithms where there is little to no dependency or requirement for communication between parallel tasks are called \textit{embarrassingly parallel}; a property that makes ray tracing an ideal candidate for execution on hardware accelerators such as GPUs.

\subsection{GPU Acceleration of Ray Tracing}

The increase in computational capacity of commodity GPUs has resulted in real time ray tracing to become pervasive across modern GPUs.
Both major GPU manufacturers, NVIDIA and AMD, offer ray tracing capabilities in off-the-shelf GPUs that typically utilize dedicated ray tracing cores in GPUs.
The main target market is gaming which has a baseline requirement the achievement of over 30 frames per second while ray tracing at 1080p or 4K resolutions.
Both AMD and NVIDIA offer Software Development Kits (SDKs) and libraries that developers can use to access the ray tracing capabilities of supporting GPUs.
These libraries and SDKs are typically implemented in CUDA~\cite{cuda:2022}, Vulkan~\cite{vulkan:2022}, OpenCL~\cite{opencl:2022}, and OpenGL~\cite{opengl:2022} and programers can utilize those via native bindings to their programming languages of choice~\cite{mamba-tracer:2022, python-rtx:2018}.

\subsubsection{Ray Tracing in Java}

The Java programming language, although not typically associated with game development, has found tremendous success in the form of the Java-based Minecraft game.
To improve graphics and utilize ray tracing, independent developers and companies have started providing ``mods'' that enable high-fidelity ray tracing.
These additions are typically developed in heterogeneous programming languages (e.g. CUDA and OpenGL), and are integrated inside the JVM via native bindings through the Java Native Interface (JNI) sacrificing performance in the process.
Since, the Java Virtual Machine, does not natively support GPU (or any from of) acceleration, the only way to enable such high-performance graphics functionality is to fragment the code base by mixing Java code with other programming languages; a fact that contradicts the ``write-once-run-everywhere'' premise of Java.

Recently, however, significant efforts have been made in order to augment the JVM with capabilities to automatically and transparently accelerate Java code on heterogeneous hardware accelerators such as GPUs, FPGAs, and others.
Prime examples of such frameworks are TornadoVM~\cite{tornadovm:2022}, Aparapi~\cite{aparapi:2022}, and IBM J9~\cite{ibmj9} which offer different degrees of functionality and JVM interoperability.
Since these frameworks enable the automatic acceleration of Java code on GPUs, a natural question that may arise is \textit{``Can we implement a complex ray tracer fully in Java and achieve real-time performance?''}; a question that this paper endeavors to answer.

\subsection{TornadoVM Background}

In this work, we utilize the TornadoVM framework~\cite{Fumero:DARHH:VEE:2019, fumero:2018} to design and build a real time GPU-accelerated ray tracer fully in Java; mainly due to its proven capabilities to achieve high performance graphics applications~\cite{10.1145/3140607.3050764}.
TornadoVM offers an API that allows developers to identify which methods and loops to parallelize through a set of annotations, such as the \textit{@Parallel} annotation.
As an example, lines 1-4 of the code snippet shown in Listing~\ref{listing:tornadovm:example} illustrate the use of the annotation on a simple method performing an element-wise addition of integer arrays \textit{a} and \textit{b} (adapted from~\cite{fumero:2018}).


\lstset{language=Java}
\begin{lstlisting}[float,label=listing:tornadovm:example,caption={Example in TornadoVM.}, xleftmargin=.05\textwidth, language=Java]
public void add(int[] a, int[] b, int[] c) {
    for (@Parallel int i = 0; i < c.length; i++) 
        c[i] = a[i] + b[i];
}
public void compute(int[] a, int[] b, int[] c) {
    TaskSchedule ts = new TaskSchedule("s0");
    ts.streamIn(a, b);
    ts.task("t0", this::add, a, b, c);
    ts.streamOut(c);
    ts.execute();
}
\end{lstlisting}

	\begin{lstlisting}[label=code:java, caption={Java snippet used to illustrate the FPGA code generation.}, xleftmargin=.05\textwidth]
public void run(float[] params ...) {
  for (@Parallel int i = 0; i < n; i++) {
    for (int j = 0; j < m; j++) {
  // computation
}}}
	\end{lstlisting}

To execute such a method with TornadoVM, instead of simply invoking the method, the user should instantiate a \texttt{TaskSchedule} object, which acts as the executor service for the routine.
A \texttt{task} can then be defined using a name, a reference to the target method as well as the data it operates on (essentially the arguments of the function).
Additionally, as hardware accelerators usually contain their own memory, the user may set up which data is transferred between the host and the hardware accelerators:

\begin{itemize}
    \item The \texttt{streamIn()} function allows the user to indicate the memory region that gets copied to the accelerator at every invocation of the task.
     Data that is required by the method but is not marked with \texttt{streamIn()} is copied to the target device at the very first call, where it persists for subsequent executions of the method.
    \item The \texttt{streamOut()} function allows the user to indicate the memory region that gets copied from the accelerator back to the host device (the main CPU).
\end{itemize}

Lines 6-10 of Listing~\ref{listing:tornadovm:example} show how a task may be composed for the previously mentioned \texttt{add} method.
Once the task, alongside its TaskSchedule is set up, the method can be invoked using an \texttt{execute()} call.
For each task, TornadoVM generates low-level code with device-specific optimizations for hardware accelerators, that perform identical computations to the generic high-level method, which is then executed on heterogeneous hardware to exploit parallel execution. 
In the next section, we describe the design and implementation of a Java ray tracer using TornadoVM.

\section{Design of the Java Ray Tracer}
\label{section:design}

This section shows the design of the proposed Java framework for real-time ray tracing processing. 
We outline the steps and algorithms involved to compute a scene using ray tracing techniques, highlighting the most compute-intensive parts of the application which are accelerated by TornadoVM. 

\subsection{Camera \& perspective}
\label{section:camera}

The \textit{camera} is the view point from which the scene is perceived, often referred to as the \textit{eye}. 
There are four major parameters that describe the camera:

\begin{centering}
\begin{tabular}{@{}>{\bfseries}l l}
 Position & The location defined by x, y, and z coordinates \\ 
 Yaw & Angle describing horizontal rotation \\
 Pitch & Angle describing vertical rotation \\
 FOV & Angular extent of a given scene that is imaged \\
\end{tabular}
\end{centering}

The position of the camera defines where the user is looking from, while the yaw and pitch values define the direction the user is looking towards. 
The field-of-view (FOV) is the angle that defines how wide the view spectrum is.
Having a single-point camera allows for a \textit{perspective} view of the scene similarly to how humans see the physical world, where objects that are further away from the observer appear smaller than those closer. 
This provides an illusion of \textit{depth}, which allows for a 2-dimensional image to gain the additional information for representing 3-dimensional objects.

\subsection{Window \& viewport}
\label{section:norm-view}

The \textit{window} is the rectangular area on the screen that displays the results of the rendering process. 
This is a physical array of pixels defined by a resolution denoting the width and the height of the window. 
A pixel on a window can be defined using \textit{screen coordinates (x, y)} (for instance, in a window of resolution $1280 x 720$, the center-most pixel has coordinates of $640 x 360$).

\begin{figure}[t!]
\centering
  \includegraphics[width=\linewidth]{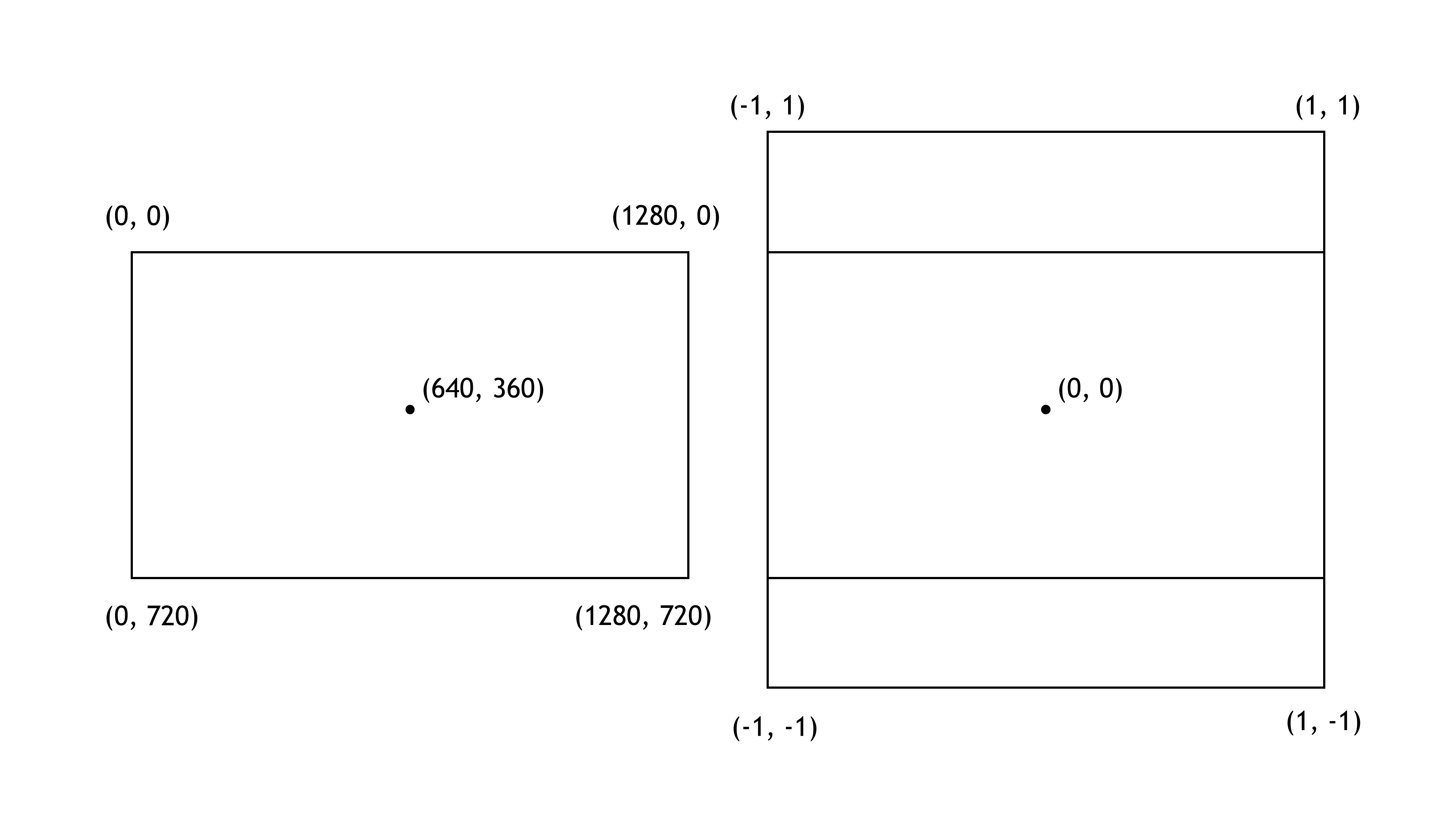}
  \caption{Mapping pixels from the window to the viewport.}
  \label{fig:norm-screen-coords}
  \vspace{-1.3em}
\end{figure}

The \textit{viewport} is the rectangular region placed in front of the camera within the Cartesian coordinate system of the scene that represents the range or area currently being viewed.
While the pixels on the window are represented by screen coordinates, the equivalent points on the viewport takes the form of \textit{normalized device coordinates (NDC)} \cite{MCREYNOLDS200519} in the range of [-1, 1], [-1, 1].
Thus, the viewport is placed inside of a square area of which each side is of size of 2 with the middle-point located at exactly $(0, 0)$ (See Figure \ref{fig:norm-screen-coords}).
The window and the viewport relate to each other with a one-to-one mapping of pixel coordinates to NDC coordinates, computed using a simple normalization function (Listing \ref{listing:getCoordinates}).

\begin{figure}[b!]
  \begin{lstlisting}[caption={Computation of a Normalized Device Coordinates for each pixel.},label={listing:getCoordinates},captionpos=b,
          xleftmargin=.023\textwidth,xrightmargin=.023\textwidth,language=Java]
getNormScreenCoords(x, y):
  if width > height:
    u = (x - width / 2 + height / 2) / height * 2 - 1;
    v =  -( y / height * 2 - 1);
  else:
    u = x / width * 2 - 1;
    v =  -((y - height / 2 + width / 2) / width * 2 - 1);
  return u, v;
\end{lstlisting}
\end{figure}

\subsubsection{Relative camera placement}
\label{section:relative-camera}

\begin{figure}[t!]
\centering
  \includegraphics[width=5.5cm]{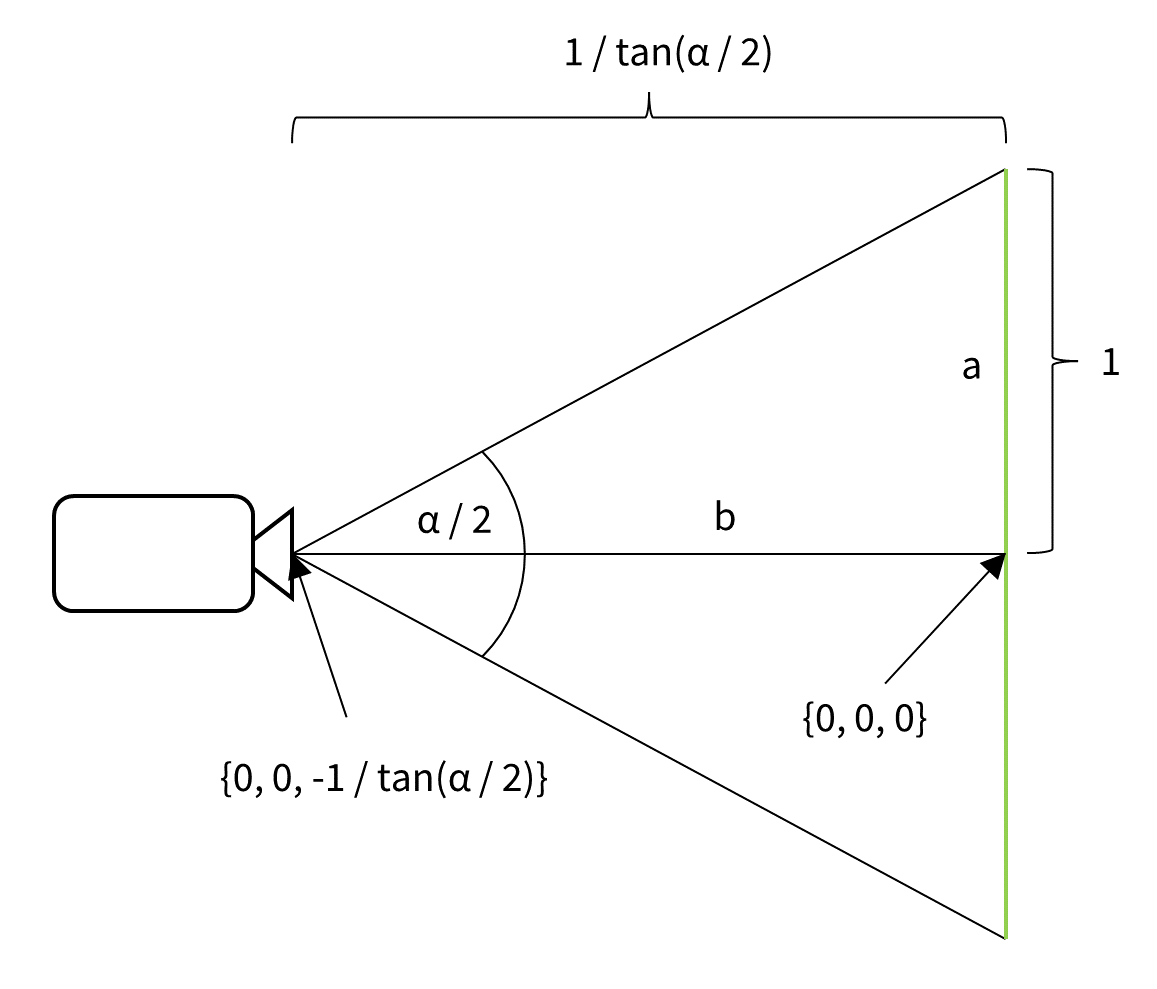}
  \caption{Camera placement relative to the viewport.}
  \label{fig:relative-camera}
  \vspace{-1.5em}
\end{figure}

To ensure that the camera view encapsulates the entire viewport, the camera is placed at a fixed distance from the viewport according to the field-of-view angle.
Figure \ref{fig:relative-camera} shows a graphical representation of the geometrical computation that allows us to define the relative position of the camera.
Note that the green line denotes the viewport.
Besides, the width of the viewport is of $2$ unit lengths, which makes half of the width denoted by $a$ equal to $1$.
Given the camera field-of-view angle $\alpha$, it can be observed that $a$ and $b$ are the opposite and adjacent sides of a right-angled triangle with an angle of $\frac{\alpha}{2}$
The distance of the camera from the viewport (denoted by $b$) can thus be computed as the quotient of $a$ and the tangent of $\frac{\alpha}{2}$:
$$
b = \frac{a}{tan(\frac{\alpha}{2})} = \frac{1}{tan(\frac{\alpha}{2})}
$$

\subsection{The primary view rays}

Once the camera and the viewport are defined, then we define the initial view rays which are shot from the camera through every pixel on the viewport into the scene. 
This will produce a total number of \texttt{width * height} rays, which are independently traced around the environment to obtain a final color for the respective pixel.

\subsubsection{Defining the rays}

Each view ray is constructed with its origin set as the position of the camera, while the direction is obtained as follows:

\begin{itemize}
    \item The viewport is placed into the center of the coordinate system in parallel to the \textit{xy plane}, with the center pixel lined up with the origin point $O(0, 0, 0)$.
    \item The normalized location of the pixel in discussion is obtained as described in section \ref{section:norm-view}. 
    For instance, the center pixel at (640, 360) in a viewport of resolution (1280, 720) will have a location of $P(0, 0, 0)$.
    \item The camera is placed at the location $C(0, 0, -1/tan(\frac{\alpha}{2}))$, where $\alpha$ is the field of view of the camera as described in section \ref{section:relative-camera}.
    \item The relative direction vector from the camera to the pixel is acquired by subtracting the location of pixel $P$ from the camera position $C$ and normalizing the result.
    \item The resulting vector is rotated around the yaw and the pitch of the camera to obtain the final direction of the ray.
\end{itemize}

\subsection{Objects in the scene}

Objects in the scene within this application take the form of primitive shapes (e.g., spheres and planes) and are defined by four properties:

\begin{centering}
\begin{tabular}{@{}>{\bfseries}l l}
 Position & x, y, z coordinates of the object's location \\ 
 Size &  size of the object, in case of spheres this \\
     &   value corresponds to the radius \\
 Color & red, green, and blue values defining color \\
 Reflectivity & value defining how reflective the \\
              & object is (explained in section \ref{paragraph:specular-lighting}) \\
\end{tabular}
\end{centering}

To illustrate the effects resulting from implementing the different features of the rendering engine (outlined in the following sections), we provide a plane and three spheres for explaining the ray tracing algorithms.



\subsubsection{Acquiring the closest intersected object}

\begin{figure}[t!]
\centering
  \includegraphics[scale=0.1]{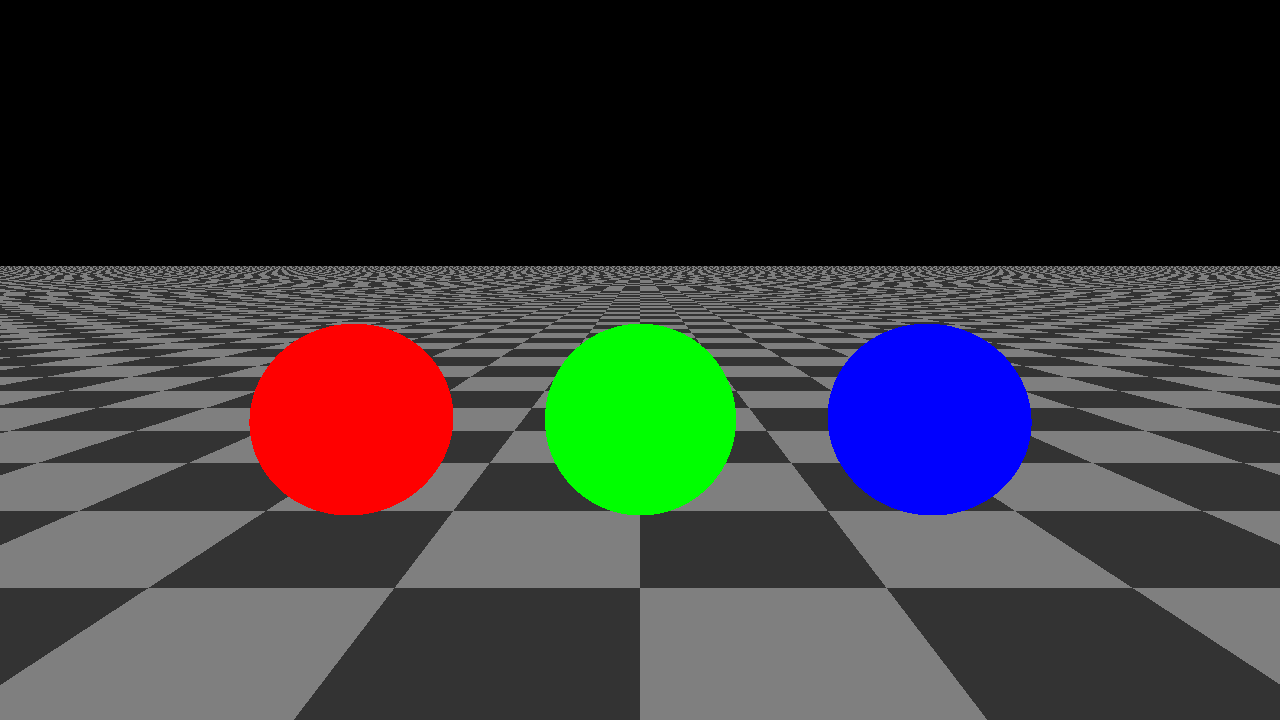}
  \caption{The result of acquiring the closest hit.}
  \label{fig:plain-colors}
  \vspace{-1.2em}
\end{figure}

Once a view ray is generated, it is then checked for intersections with every object in the scene and returns the index and the intersection point on the object that is the closest to the origin of the ray.
In cases where no intersections are found, a plain black background color is returned, which is later replaced by a skybox image (Section \ref{section:additional-components}).
By defining and tracing the primary view rays to acquire the closest intersected object and simply returning the base color, the output shown in Figure \ref{fig:plain-colors} is obtained.

\subsection{Basic Shading}

The following step in the implementation of the rendering engine entails an addition of a light source which provides some local illumination to the scene.
Knowing the location of the light source, the direction from which light rays come from can be defined, allowing for shading effects to be computed, providing a 3-dimensional look to objects instead of the flat appearances we observe in Figure \ref{fig:plain-colors}.

\subsubsection{The Phong illumination model}

\begin{figure}[b!]
\centering
  \includegraphics[width=\linewidth]{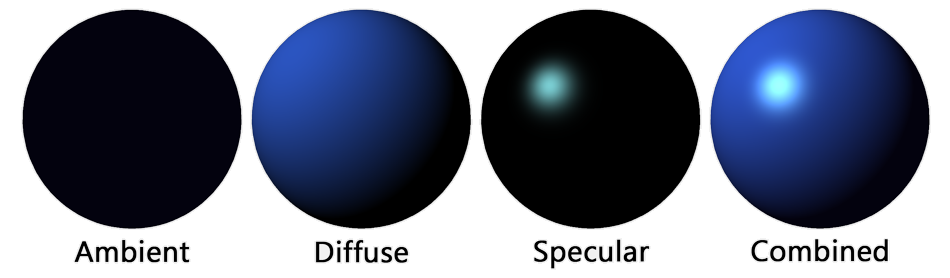}
  \caption{The components of the Phong illumination model.}
  \label{fig:am-diff-spec}
  \vspace{-1.2em}
\end{figure}

To achieve high realism in our scene, we employ empirical models of illumination to perform shading, which are based on real-life observations of light behavior. 
An efficient and realistic model widely used today in computer graphics was described by Bui Tuong Phong in 1975 \cite{phong:1975}, which was also utilized in the paper of Turner Whitted \cite{whitted:1980}.
The Phong illumination model combines three main components that contribute to a final color: ambient, diffuse, and specular, as illustrated in Figure \ref{fig:am-diff-spec}.

\vspace{-0.2cm}\paragraph{Ambient lighting}

In a real-life scenario, light would not originate from one single source, as there might be additional illumination such as distant sunlight or moonlight that contribute to the image even if they are not directly in sight. Shadows are thus never entirely pitch black. To simulate the resulting effect, a constant ambient strength value in the range of [0, 1] is described as a small percentage of the object color that is always visible, regardless of whether the point is on a shaded side or not.
In ray tracing, this component is especially crucial as rays would have to be traced towards every single point that emits light within an environment to compute colors otherwise.


\vspace{-0.2cm}\paragraph{Diffuse lighting}
\label{paragraph:diffuse-lighting}

The diffuse component models the directional impact of light on an object, based on the simple principle that the more a surface faces away from a light source, the darker it appears. This is a result of the object itself blocking light from arriving at these surfaces.
The component is defined through the dot product (the angular distance) between the normal vector of a surface point (the direction the surface is facing), and the direction vector traced from the surface point to the light (the direction the light rays are coming from).
Diffuse lighting~\cite{goldstein:1971} is the most significant component of the Phong illumination model that gives objects a 3-dimensional shaded appearance by producing shadows on surfaces facing away from the light, commonly referred to as \textit{form shadows}. 




\vspace{-0.2cm}\paragraph{Specular lighting}
\label{paragraph:specular-lighting}

\begin{figure}[t!]
\centering
  \includegraphics[scale=0.2]{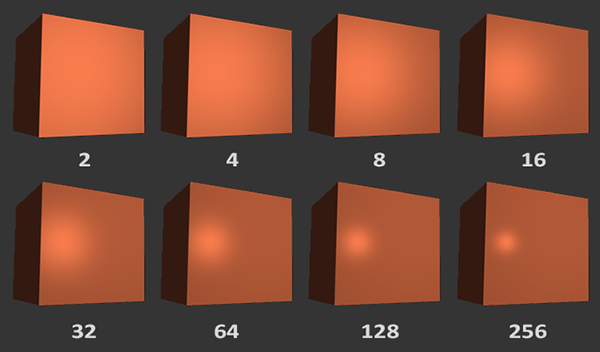}
  \caption{The different levels of reflectivity values.}
  \label{fig:reflectivities}
\end{figure}

The specular component displays a bright spot on the surfaces on shiny objects as a result of looking at reflected light beams. 
The more reflective a surface, the smaller and more concentrated the spot is, whereas on dull surfaces, it tends to spread over a larger area.
This is where a \textit{reflectivity} value is defined for each object: an exponent value which defines how shiny/reflective the given object is and thus determines the spread of the specular highlight. 
The appearance of the spot at different levels of the reflectivity value is shown in Figure \ref{fig:reflectivities}.
The specular highlights are calculated using a reflection vector, which is obtained through reflecting the light direction around the normal vector of the respective surface point.
The angle between that reflection vector and the direction to the camera defines a specular factor which when combined with the reflectivity value of the object yields the strength of the highlight.
Listing~\ref{listing:diffuse} shows the corresponding pseudocode.
The result is a 
value obtained per pixel to enable an effect similar to the right-hand side of Figure~\ref{fig:am-diff-spec}.

\begin{figure}[t!]
\begin{lstlisting}[caption={Pseudocode for the computation of the Phong illumination model.},label={listing:diffuse},captionpos=b,
          xleftmargin=.023\textwidth,xrightmargin=.023\textwidth,language=Java]
normal = normalAt(surfacePos);
viewDir = normalize(cameraPos - surfacePos);
lightDir = normalize(surfacePos - lightPos);
// Reflection vector
reflectDir = reflect(lightDir, normal);
// Specular strength
specular = max(dot(viewDir, reflectDir), 0.0);
// Reflectivity is an exponent that defines the spread
specular = pow(specular, objectReflectivity);
// Highlight takes the color of the light
result = specular * lightColor;
\end{lstlisting}
\end{figure}

\subsubsection{Blinn's revision of specular highlights}

\begin{figure}[b!]
\centering
  \includegraphics[width=\linewidth]{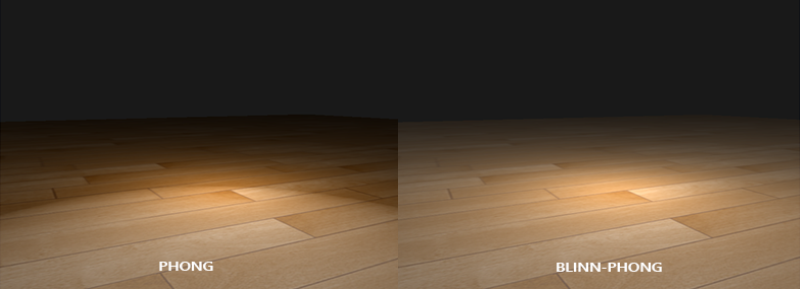}
  \caption{The cutoff from Phong's specular highlights.}
  \label{fig:phong-problem}
  \vspace{-1.6em}
\end{figure}

Acquiring the strength of the specular highlight from the angle between the reflection vector and the view direction contains a small flaw: 
if a surface point is further away from the light source, then the reflection vector ends up pointing far away into the distance.
This entails that a set points on object surfaces that are behind the light with respect to the camera, produces reflection vectors with angles larger than 90° degrees to the view direction. 
In these instances the dot products evaluate to be negative, resulting in no specular highlights. This cutoff is illustrated in Figure \ref{fig:phong-problem}.

\begin{figure}[ht] 
  \begin{subfigure}[b]{0.5\linewidth}
    \includegraphics[width=0.95\linewidth]{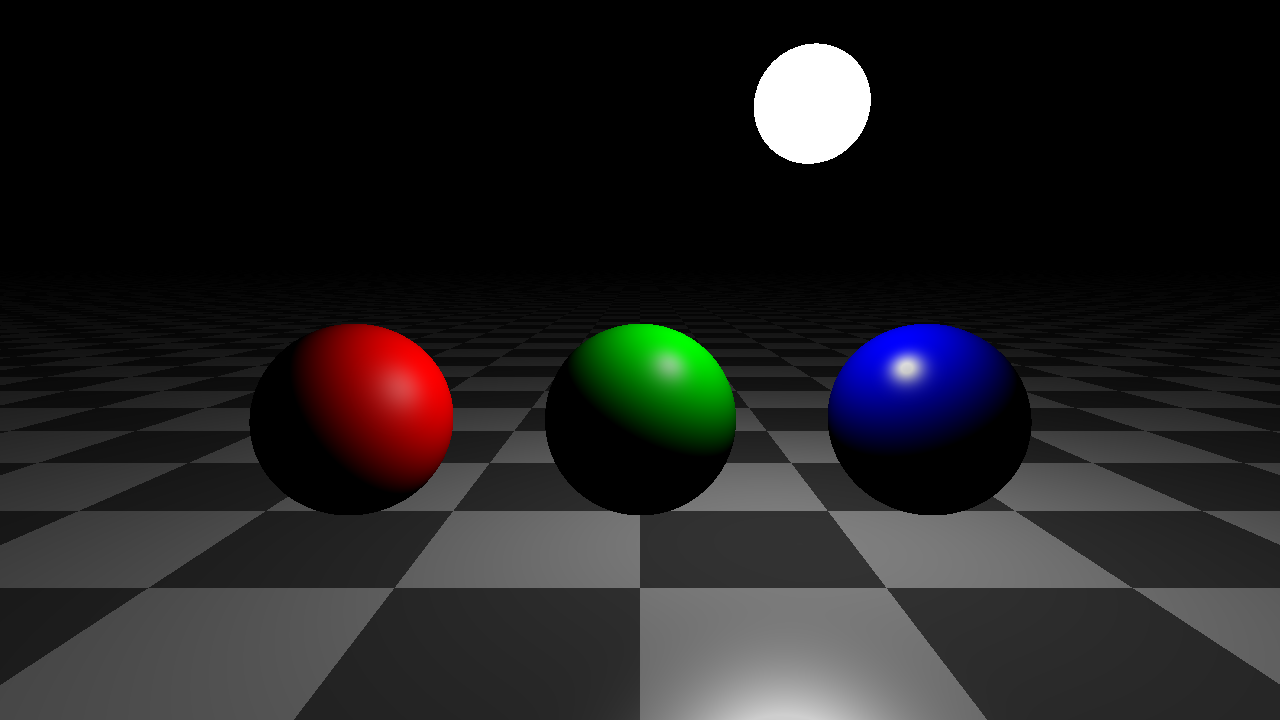}
  \caption{Application of the Blinn-Phong shading model.}
  \label{fig:blinn-phong}
  \end{subfigure}
  \begin{subfigure}[b]{0.5\linewidth}
    \includegraphics[width=0.95\linewidth]{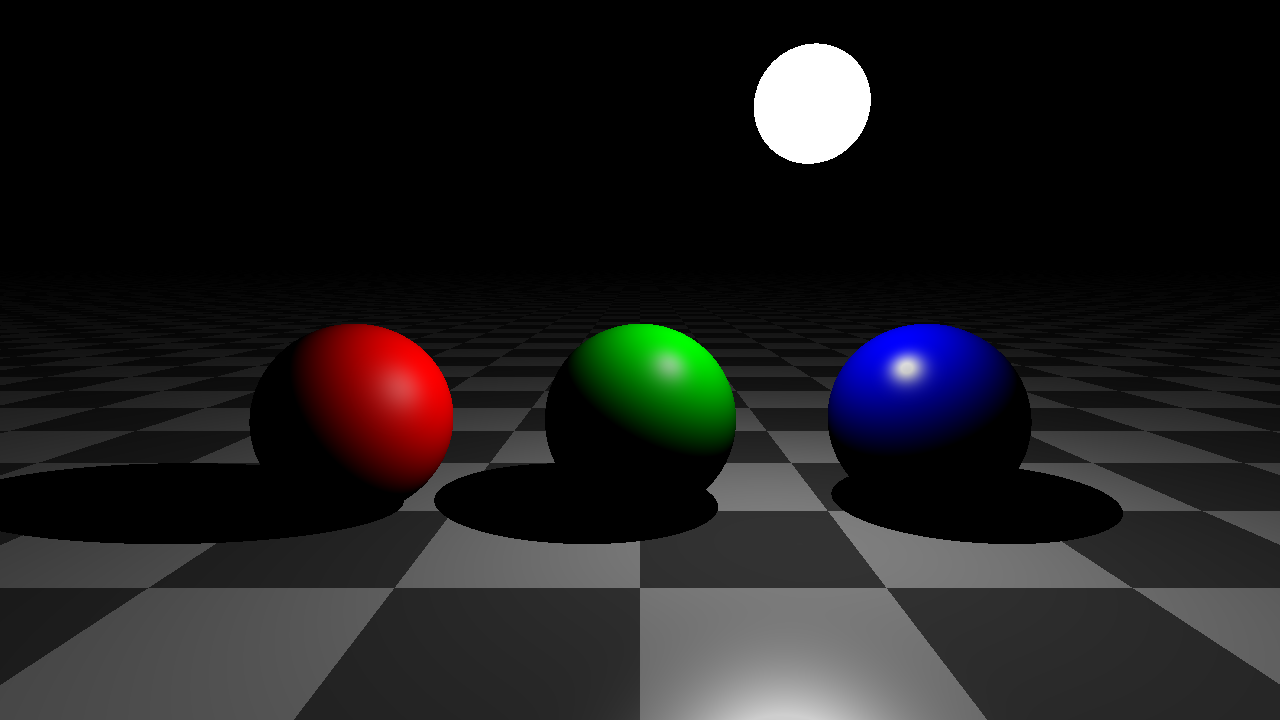}
    \caption{Addition of cast shadows.}
    \label{fig:hard-shadows}
  \end{subfigure}
  \caption{The result of applying the Blinn-Phong shading model and adding cast shadows.}\label{models_fig}
\end{figure}

To solve this issue, James Blinn designed a modification to the Phong specular highlights in 1977 \cite{blinn:1977},
proposing to instead take the dot product between the normal of the surface and the halfway vector between the light and the view directions to keep the angle under 90° degrees.
This allows the specular-highlights to spread correctly along surfaces.
Thus, the final result of applying all three components of the combined Blinn-Phong illumination model to the objects in the scene, 
is shown in Figure \ref{fig:blinn-phong}.




\subsection{Cast shadows}
\label{section:cast-shadows}

While the Blinn-Phong illumination model has allowed for form shadows to be computed for every individual object in the scene, surfaces may also experience shading resulting from other objects blocking the light, producing \textit{cast shadows}~\cite{appel:1968}.
The accuracy
 of cast shadows is one of the main strengths of ray tracing: By generating a \textit{shadow ray} from a surface point to the light source, the ray can be checked for an intersection with the remaining objects to evaluate whether anything is occluding the surface from being lit.

When a surface point is evaluated to be under the effects of a cast shadow, the color of the surface is multiplied with the ambient strength of the scene, resulting in a darker area. Figure \ref{fig:hard-shadows} displays the result of computing cast shadows. Listing~\ref{listing:castshadows} shows a code snippet to identify if an object is in a cast shadow.

\begin{figure}[t!]
  \begin{lstlisting}[caption={Pseudocode to process cast shadows in a scene.},label={listing:castshadows},captionpos=b,
          xleftmargin=.023\textwidth,xrightmargin=.023\textwidth,language=Java]
// Generate a shadow ray
shadowRay.dir = normalize(lightPos - surfacePos);
shadowRay.origin = surfacePos;

// Acquire the distance to the light
lightDistance = distance(lightPos, surfacePos);

// Check if any objects are between the surface 
// and the light
for each object:
    if intersects(shadowRay, object) &&
       distance(object, surfacePos) < lightDistance:
        surfaceIsInCastShadow = true;
        break;
        
surfaceIsInCastShadow = false;
\end{lstlisting}
\vspace{-1.0em}
\end{figure}

\subsubsection{Soft shadows}
\label{section:soft-shadows}

\begin{figure}[b!]
\centering
  \includegraphics[width=\linewidth]{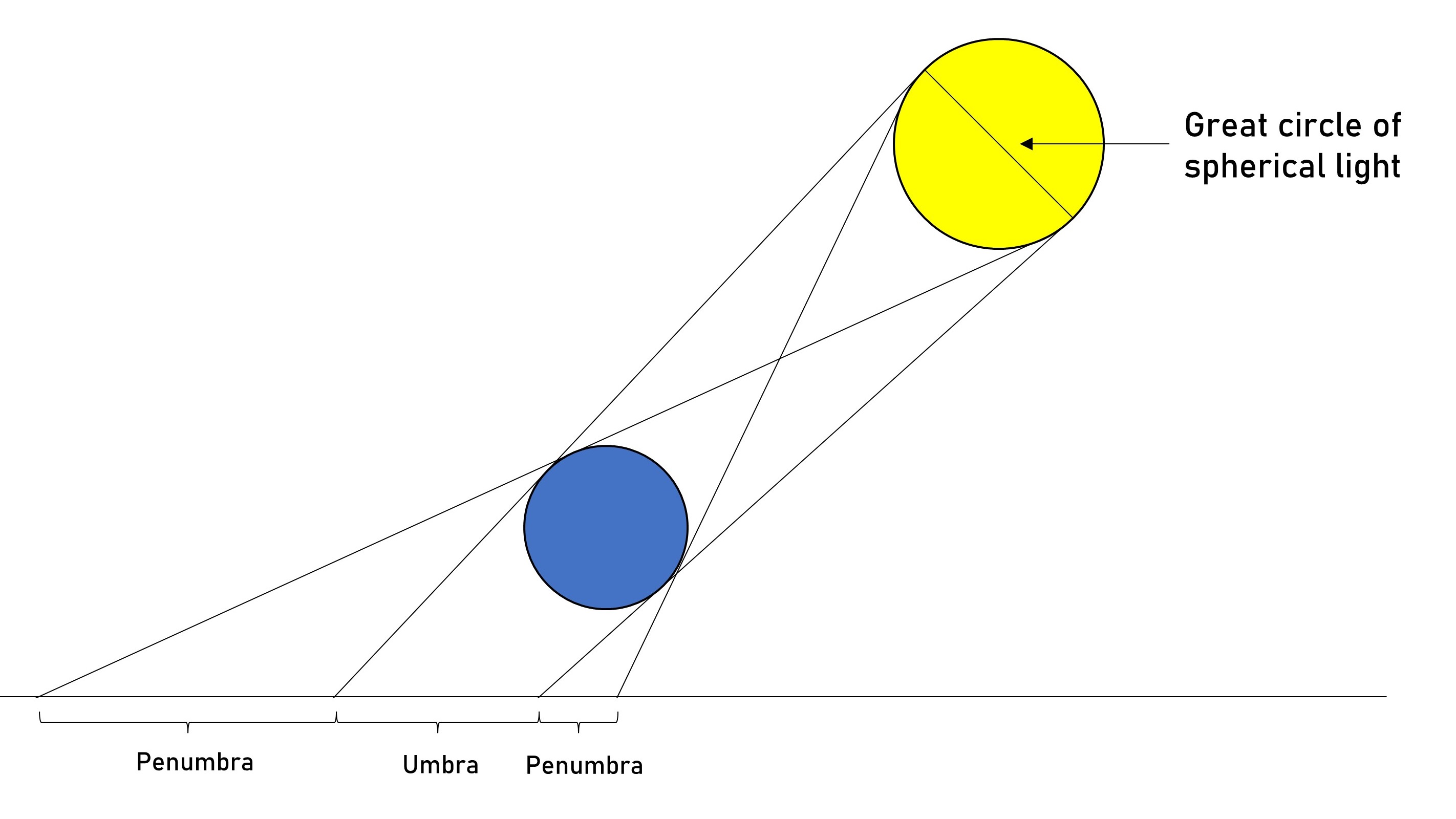}
  \caption{Illustration of umbra and penumbra shadows.}
  \label{fig:penumbra}
\end{figure}

One noticeable issue that can be observed in Figure \ref{fig:hard-shadows} is the crisp, hard edges of the cast shadows, which appear inconsistent with the fading effect that diffuse lighting has produced on the form shadows. 
This issue has led us to take a small step away from classical Whitted-style ray tracing to produce \textit{soft shadows} using a stochastic sampling method.
In real life, instead of coming from a single point, light rays originate from an area encompassing the volume of an emitter. 
The inner region of a surface in shadow, where every light ray coming from the light source is blocked and is fully shaded as a result, is called the \textit{umbra}, while the transitional region that gets lighter towards the edge due to less and less light rays being obstructed, is called the \textit{penumbra} (See Figure \ref{fig:penumbra}).
The phenomenon produces a softer and more realistic look to the cast shadows \cite{soft-shadows:1984}.


In the case of a spherical light, the area where light rays originate from can be modelled as the great circle facing the surface point in question. 
The effects of soft shadows are simulated by generating an \textit{n} number of sample points on this circle and tracing a shadow ray to each point. 
The number of rays that are blocked by another object divided by the total number of sampled rays defines the brightness of the given surface point. 
This results in a shadow coefficient in the range of [0, 1] to be multiplied with a given surface point to darken the respective area.

\paragraph{Uniform sampling of points on a circle}

\begin{figure}[b!]
\centering
  \includegraphics[width=6cm]{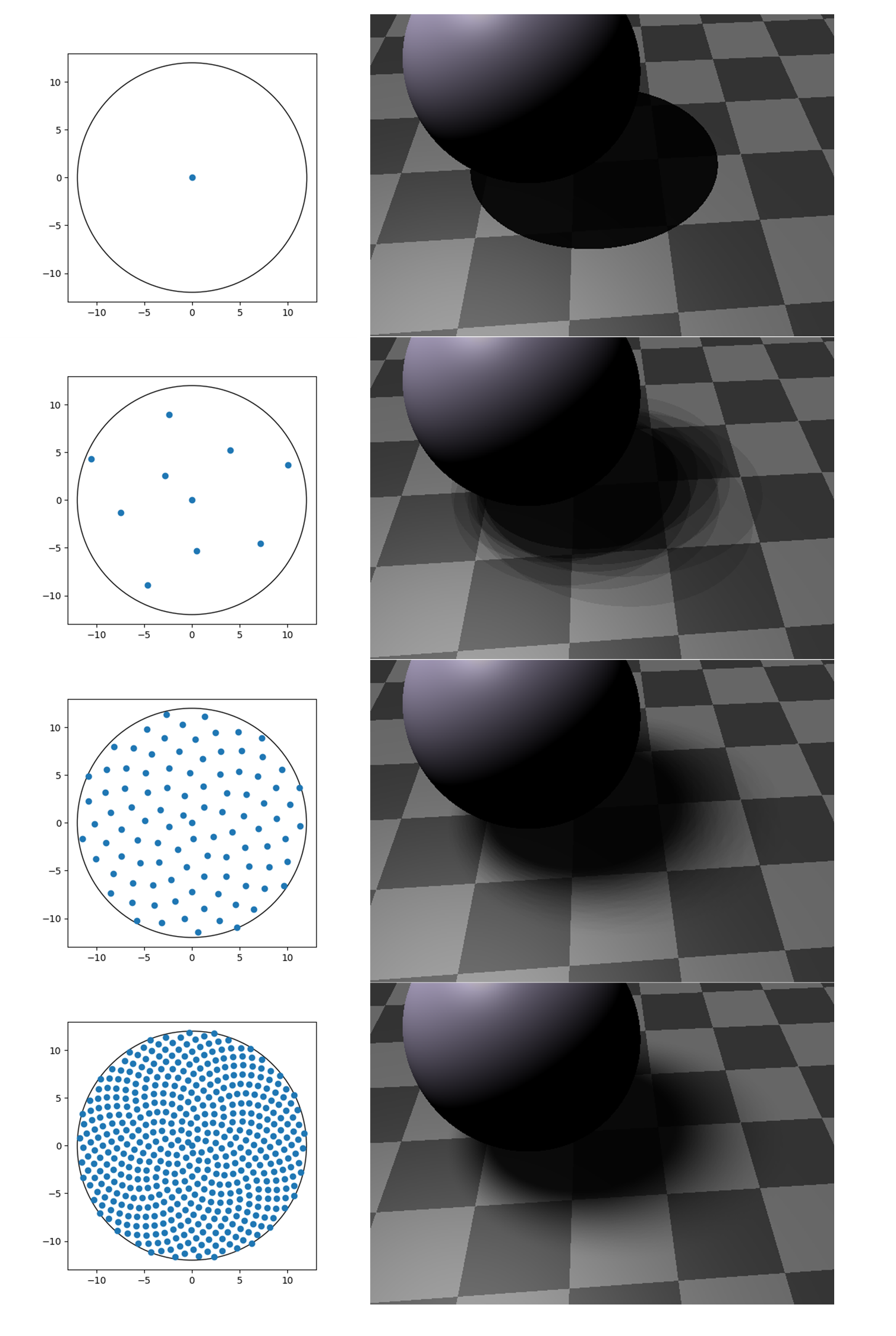}
  \caption{Light sample sizes from top to bottom: 1, 10, 100, 500.}
  \label{fig:sampling}
\end{figure}

\begin{figure}[t!]
\centering
  \includegraphics[width=0.65\linewidth]{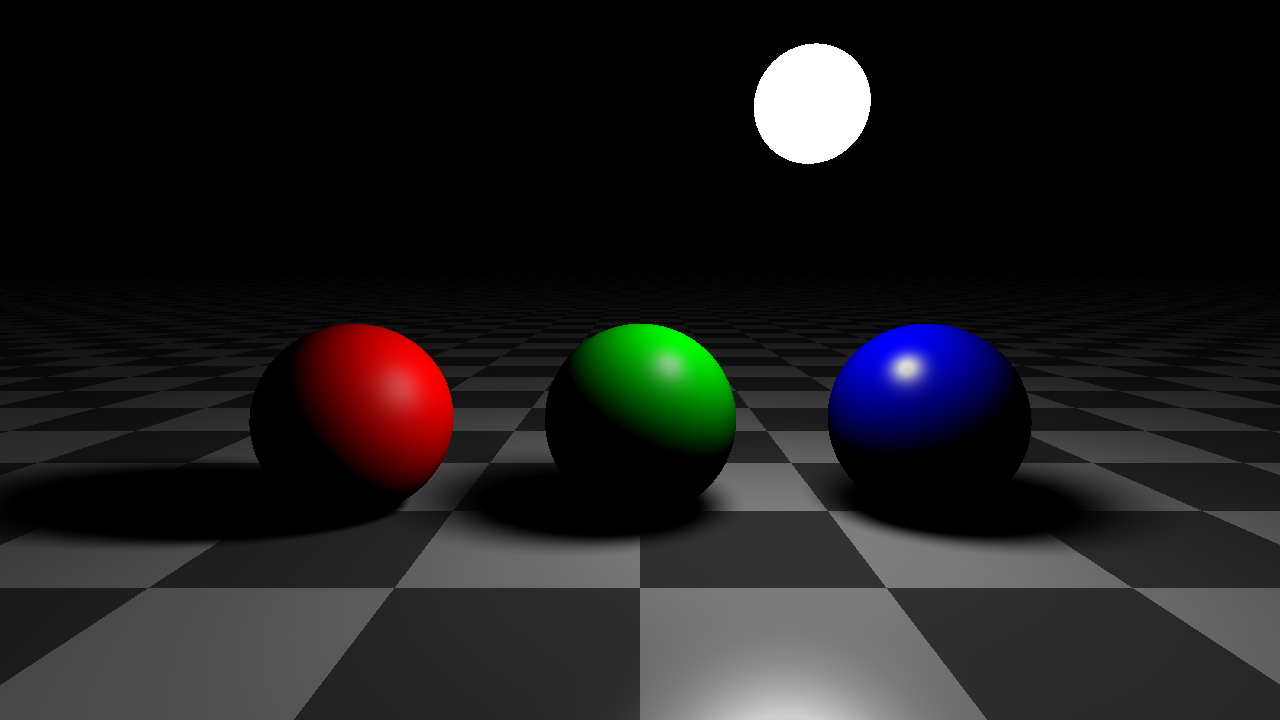}
  \caption{The result of sampling the cast shadows.}
  \label{fig:soft-shadows}
  \vspace{-1.0em}  
\end{figure}

To avoid cluttering of points during sampling, obtaining an \textit{n} number of samples that cover an area with the most equal scattering possible, requires a form of uniform distribution.
The algorithm utilized in this paper is based on a method known as the \textit{sunflower seed arrangement}~\cite{vogel:1979}.
The model generates each point \textit{i} in a total of \textit{n} in the following manner:

$$
r(i) = 2 \times radius \times \sqrt{\frac{i}{n}} \\
\theta(i) = i \times \phi
$$
where \textit{r(i)} is the distance from the circle's origin and \textit{$\theta(i)$} is the angle at which point \textit{i} resides, with \textit{$\phi = \pi \times (3 - \sqrt{5}) \text{ rad} \approx 137.507\degree$} being the golden angle.
Figure \ref{fig:sampling} illustrates the effects of generating different numbers of sample points, notice the improvement in quality as the sampling size increases.
The final result of computing soft cast shadows with a sample size of 300 is shown in Figure \ref{fig:soft-shadows}.

\subsection{Reflections}
\label{section:reflections}

\begin{figure}[t!]
\centering
  \includegraphics[width=\linewidth]{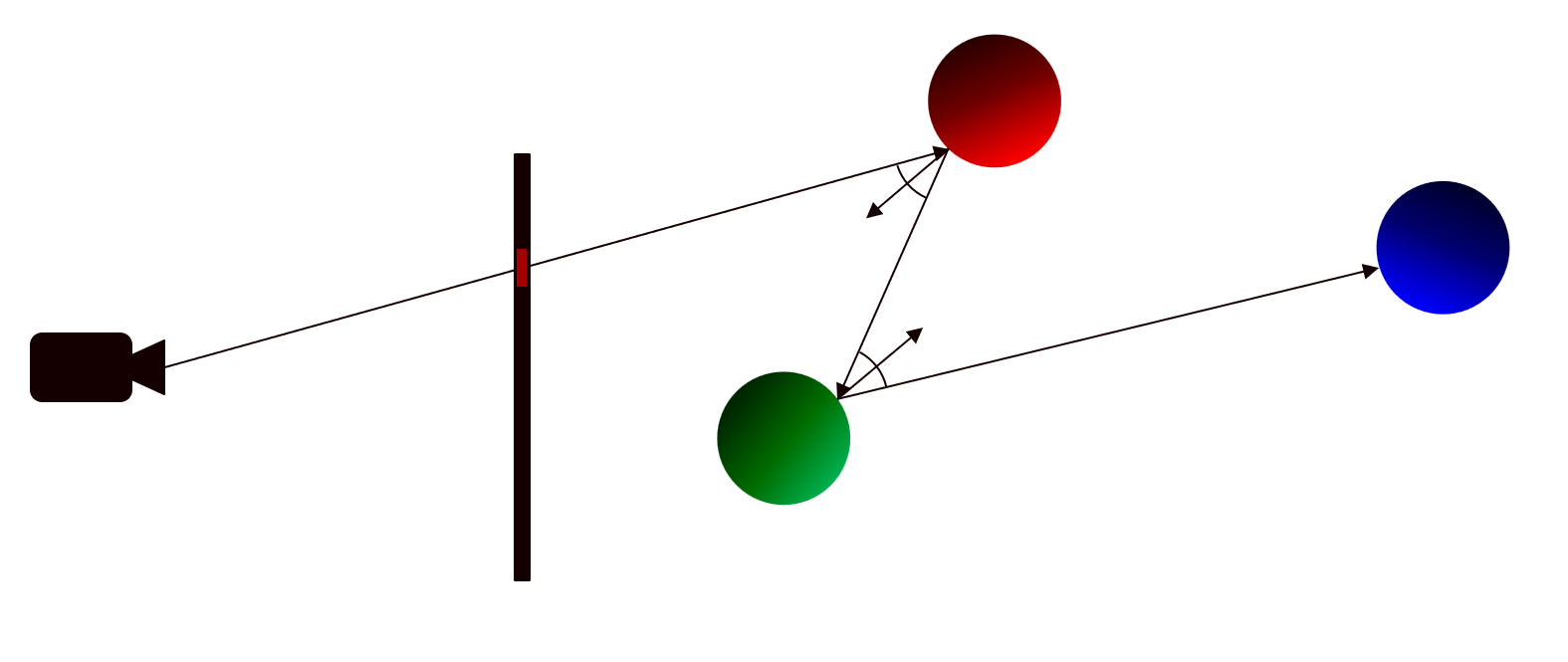}
  \caption{Generating and tracing reflection rays.}
  \label{fig:refl-rays}
\end{figure}

With shading and shadows in place, the final step of the rendering engine entails the addition of color transport between scene objects through reflections, encompassing the works of John Turner Whitted \cite{whitted:1980}.
To compute reflections, when a view ray finds its first intersected object, next to the computation of shading effects and casting shadows to determine a final color value, an additional \textit{reflection ray} is spawned with a direction symmetrical to the incoming ray's direction (Snell's law of reflection \cite{snell:2017}).
The ray is then traced throughout the scene in the same manner as the initial view ray to acquire a reflection color (Figure \ref{fig:refl-rays}).
To obtain a final color, the base color of the object is mixed with the reflection color using a ratio depending on how reflective the object is. This ratio in our case, is derived from the reflectivity value introduced during the computation of specular highlights (Section \ref{paragraph:specular-lighting}), by dividing the reflectivity exponent with a predefined constant \textit{MAX\_REFLECTIVITY} value (objects that have a reflectivity equal to \textit{MAX\_REFLECTIVITY} is a fully reflective mirror-like surface).


\subsubsection{Recursive reflections}

The aforementioned process, however, does not have to halt once the reflection ray has gathered color information at its first hit. 
In this event, another subsequent reflection ray may be sent out, continuing the algorithm in a recursive manner to produce reflections within reflections (Figure \ref{fig:refl-rays}).
To avoid a potentially infinite amount of reflections rays generated, a maximum depth is defined to limit the recursion to a specific number of ray bounces.
Listing~\ref{listing:reflections} shows a pseudocode for computing the color of a trace with depth $N$. 

\begin{figure}[t!]
  \begin{lstlisting}[caption={Color calculation within a trace of specific depth.},label={listing:reflections},captionpos=b,
          xleftmargin=.023\textwidth,xrightmargin=.023\textwidth,language=Java]
// Trace the ray into the scene to obtain color
rayTrace(ray, depth):
  // Acquire the closest intersected object as well as
  // the point of intersection
  object, hitPos = getClosestHit(ray);
  // If ray doesn't hit any objects, then return black
  if object == null:
      return BLACK
  else:
    // If reflection bounce limit is reached then
    // perform Blinn-Phong shading and cast shadows
    // to obtain color
    if depth == 0:
      return shade(object.color);      
      // Else trace a reflection ray through the scene
    else:
      normal = normalAt(hitPos);
      reflRay.dir = reflect(ray.dir, normal);
      reflRay.origin = hitPos;
      reflColor = rayTrace(reflRay, depth - 1);
      reflRatio = object.reflectivity / MAX_REFL
      // mix base and reflection color
      color = object.color * (1 - reflRatio) +
              reflColor * reflRatio;
      return shade(color);
\end{lstlisting}
\end{figure}

\begin{figure}[b!]
\centering
  \includegraphics[width=0.75\linewidth]{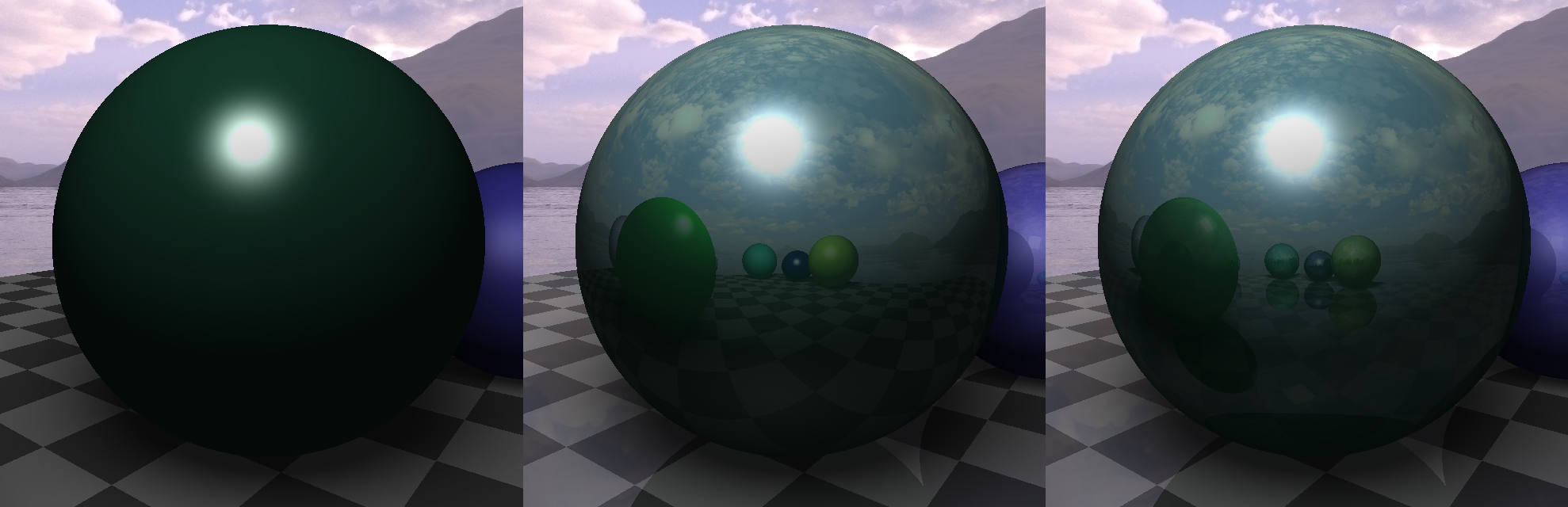}
  \caption{Limit of 0 (left), 1 (middle), 2 (right) reflection bounces.}
  \label{fig:reflection-bounces}
\end{figure}

\begin{figure}[t!]
\centering
  \includegraphics[width=0.65\linewidth]{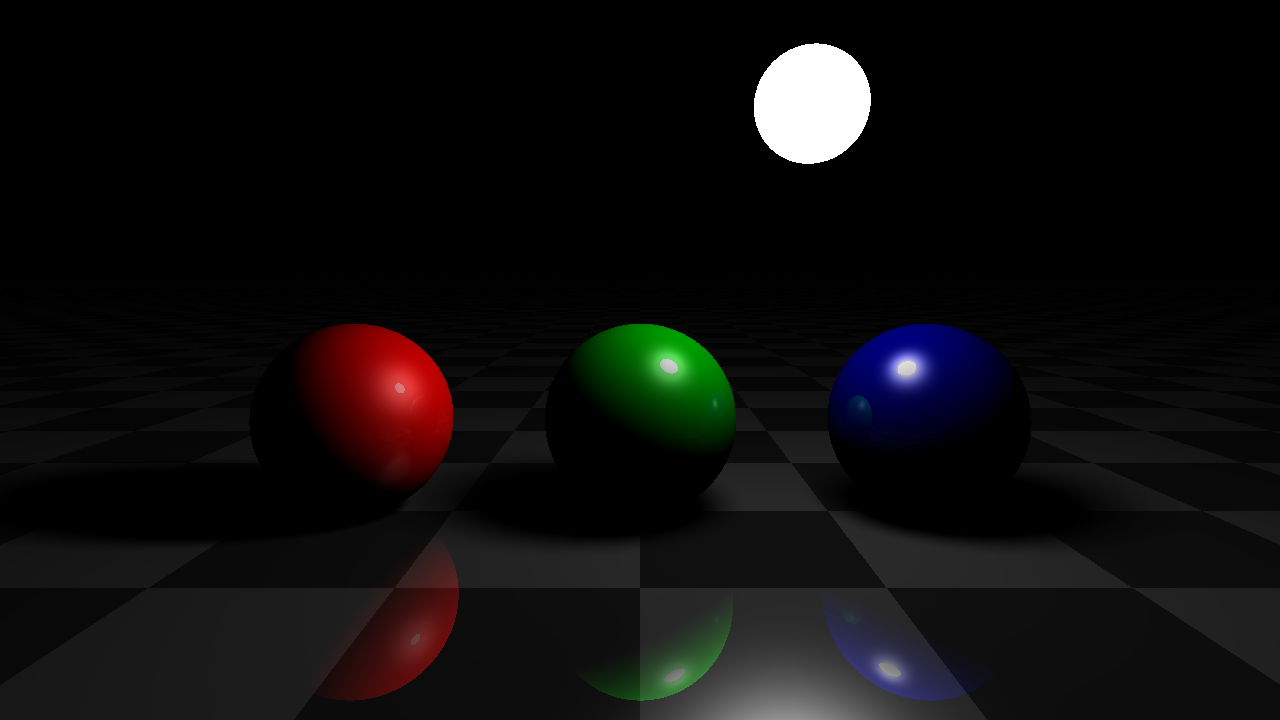}
  \caption{The result of calculating reflections.}
  \label{fig:reflections}
\end{figure}

Figure \ref{fig:reflection-bounces} illustrates the different results obtained at different recursion limits. 
We can notice that with two reflection bounces, reflections can be seen within the reflection of the sphere.
Figure \ref{fig:reflections} shows the results of calculating reflections in the basic scene within our ray tracing framework. 

\subsubsection{Iterative reflections}

While using recursion to compute reflections is intuitive, unfortunately TornadoVM does not support recursive calls due it being forbidden in the back end languages such as OpenCL and CUDA. 
Thus as a workaround, recursive reflections were emulated using a manual stack to record color as well as shading and casting shadow information at each reflection bounce, while looping through the actions of each of the reflection rays with an iterative loop.
When the reflection bounce limit is reached or a reflection ray exits the scene without an intersection, the stack is read in reverse to mix the recorded colors.

\subsection{Additional components}
\label{section:additional-components}

A number of additional components have been added to the implemented Java ray tracer including:
\begin{itemize}
\item \textbf{Skybox:} Distant background in the form of a High Dynamic Range (HDR) image for improving colors and reflectivity using UV mapping~\cite{wiki:UV_mapping}.
\item \textbf{Physics system:}  A physics system that employs the \textit{The Verlet integration}~\cite{verlet} for adding motion to rendered physical objects within a scene.

\end{itemize}

\subsection{TornadoVM Integration}


\begin{figure}[t!]
    \begin{lstlisting}[caption={Method signature for the main render method to be accelerated with TornadoVM.},label={listing:renderSignature},captionpos=b,
            xleftmargin=.023\textwidth,xrightmargin=.023\textwidth,language=Java]
Renderer::render
    (pixels:              int[],
    dimensions:           int[],
    camera:               float[],
    rayTracingProperties: int[],
    bodyPositions:        VectorFloat4,
    bodySizes:            VectorFloat,
    bodyColors:           VectorFloat4,
    bodyReflectivities:   VectorFloat,
    skybox:               VectorFloat4,
    skyboxDimensions:     int[]) : void;
\end{lstlisting}
\end{figure}

The entire rendering process described in this section is implemented within a method named \textit{render} (shown in Listing~\ref{listing:renderSignature}) with the following input arguments:

\begin{itemize}
  \item \textbf{pixels}: an array of size \texttt{width * height} where the computed color values for every pixel are stored upon executing the method.
  \item \textbf{dimensions}: a two-element array that holds the width and height of the viewport.
  \item \textbf{camera}: a float array of six elements containing the parameters of the camera such as the $x,y,z$ coordinates, yaw and pitch angle of rotation, and field of view.
  \item \textbf{rayTracingProperties}: an array of size 2, which includes the desired soft shadow sampling size and the desired limit of reflection bounces.
  \item \textbf{bodyPositions, bodySizes, bodyColors and bodyReflectivities}: These four arguments encode the information about all of the objects in the scene; each object is defined using its position, size, color, and reflectivity.
  \item \textbf{skybox}: an array of four elements that includes the color values of every pixel of the HDRI skybox image. This is used to display the panoramic background instead of a plain color. 
\end{itemize}

These arguments provide the \textit{render} method with all the required information to produce a final color for every pixel.
As the process of acquiring a final color for each pixel is independent from one another, 
the main loop iterating over the pixels within the \textit{render} method can be marked with the \textit{@Parallel} annotation~\cite{fumero:2018}.
This instructs the TornadoVM compiler to parallelize the for loop and optimize the code depending on the target architecture (in our case a GPU).
Listing~\ref{listing:render} shows a code snippet that represents the main render method expressed in the TornadoVM API using the \textit{@Parallel} annotation. 
Since there are two loops to be parallelized, the TornadoVM compiler will generate an optimized 2D kernel. 

\begin{figure}[t!]
    \begin{lstlisting}[caption={Main render Java method expressed in TornadoVM.},label={listing:render},captionpos=b,
            xleftmargin=.023\textwidth,xrightmargin=.023\textwidth,language=Java]
Renderer::render(int[] pixels, int[] dimensions, ...args...) {
    int width = dimensions[0];
    int height = dimensions[1];
    ...
    // Parallel marked loop
    for (@Parallel int x = 0; x < width; x++) {
        for (@Parallel int y = 0; y < height; y++) {
        
            // Compute color
            ray = computeViewRay(x, y);
            color = rayTrace(ray, depth);
            
            // Write the color value to output
            pixels[x + y * width] = toInt(color);
        }
    }
}
\end{lstlisting}
\end{figure}

To execute the code, a \textit{TaskSchedule} is defined as shown in Listing~\ref{listing:renderTasks}.
As shown, the arguments \texttt{camera}, \texttt{rayTracing}-\texttt{Properties} and \texttt{bodyPositions} are passed to the \texttt{streamIn()} operator to facilitate the user changing viewpoints and rendering parameters as well as the position of each object to allow for physics to move objects around in the scene.
On the other hand, the \texttt{pixels} argument is passed with \texttt{streamOut()} operator in order for the results of the render method to be copied back to the host device and its memory for display.

\begin{figure}[b!]
    \begin{lstlisting}[caption={TaskSchedule composition for accelerating the Java render method with TornadoVM.},label={listing:renderTasks},captionpos=b,
            xleftmargin=.023\textwidth,xrightmargin=.023\textwidth,language=Java]
TaskSchedule ts = new TaskSchedule("s0");
ts.streamIn(camera, rayTracingProperties, bodyPositions);
ts.task("t0", Renderer::render, pixels,
        dimensions, camera, rayTracingProperties, bodyPositions, bodySizes,
        bodyColors, bodyReflectivities, skybox, skyboxDimensions);
ts.streamOut(pixels);
\end{lstlisting}
\end{figure}

The ray tracing application has been entirely written in Java and the JavaFX framework~\cite{javafx} using \texttt{Canvas} and \texttt{Pixel}-\texttt{Writer}.
The \textit{JavaFX Canvas} is an image area defined by \textit{width} and \textit{height} that can be used for drawing, while the \textit{JavaFX PixelWriter} is an interface that defines methods which allow for pixel data to be displayed on the canvas.
The JavaFX GUI updates the UI elements at fixed intervals, usually at the refresh rate of the user's monitor screen, which allows for the interface to respond to user input, producing an interactive experience.
A feature called \texttt{AnimationTimer} is provided with the JavaFX suite, which allows the user to define a subroutine inside a handle method that gets called at every update; frequently used to create animations and game loops.
We make use of this feature to invoke TornadoVM and render the canvas with the updated pixels.

\section{Evaluation}
\label{section:evaluation}

We performed the performance evaluation by utilizing the OpenCL backend of TornadoVM which has coverage for NVIDIA GPUs, Intel integrated GPUs, and Intel CPUs among other devices.
The performance evaluation uses the following two baselines: a) a sequential execution of the Java code, and b) a parallel implementation using Java Parallel Streams.
The accelerated versions of the ray tracer have been run using the OpenCL runtime on a multi-core CPU, an integrated GPU, and a dedicated GPU.

\begin{figure}[t!]
\centering
\includegraphics[width=0.4\textwidth]{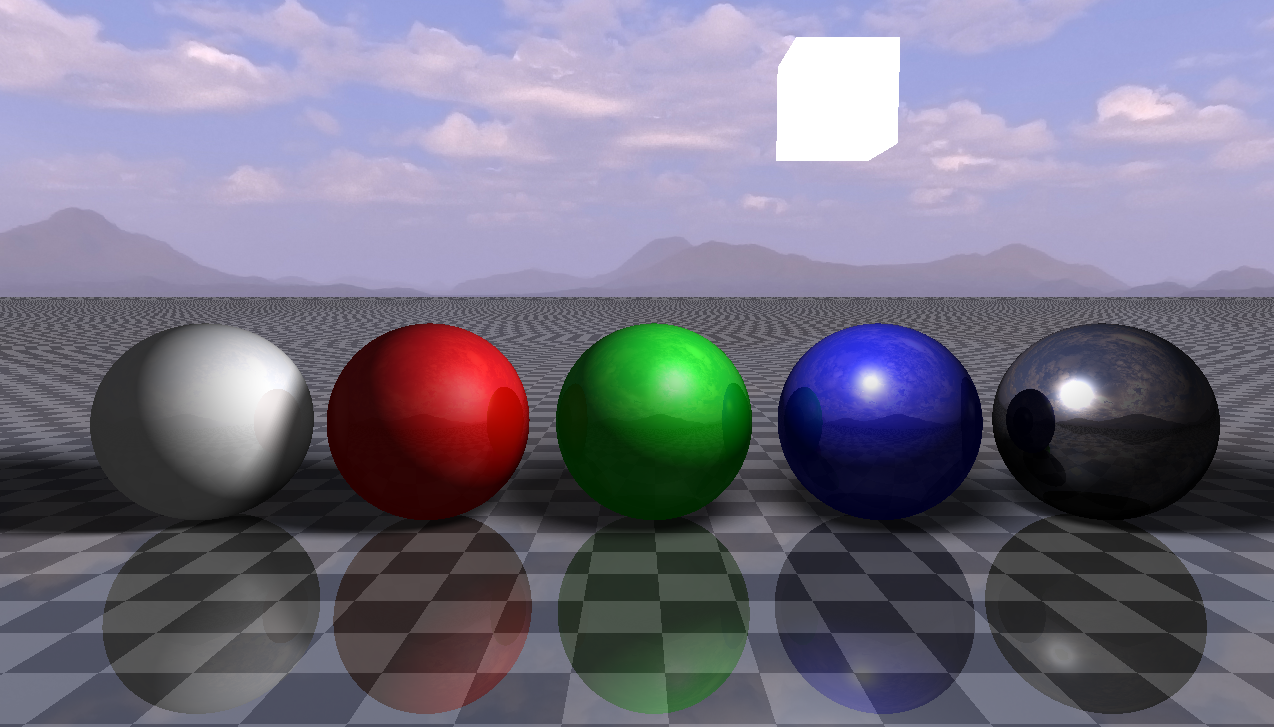}
\caption{Benchmark scene with a skybox~\cite{cgaxis:2022}.}
\label{fig:adding-skybox}
\end{figure}

\subsection{Evaluation Setup}

\begin{table}[t!]
	\centering
	\caption{The experimental hardware and software specifications of the system.}
	\label{table::platform_specs}
	\begin{tabular}{|l|c|}
		\hline
    \textbf{Operating System} & Arch Linux x86\_64 \\
    \textbf{Desktop Environment} & Gnome 42.0 \\
    \textbf{Host Device} & XPS 13 9370 \\
    \textbf{CPU} & Intel i7-8550U @ 4.0Ghz \\
    \textbf{Integrated GPU} & Intel UHD Graphics 620 \\
    \textbf{Dedicated GPU} & NVIDIA GeForce RTX 2060 \\
    \textbf{JDK} & graalvm-ce-java11-21.3.0 \\
    \textbf{TornadoVM Version} & 0.14-dev \\
		\textbf{TornadoVM Backends} & OpenCL \& PTX \\ \hline
	\end{tabular}
\end{table}

The specifications of the system used to run the tests is shown in Table~\ref{table::platform_specs}.
Regarding the evaluation methodology, we performed a warm-up phase of 100 frames when running in benchmark mode (terminal), and after 2 minutes of execution when running the GUI that renders the scene in real time. 
After the warmup phase, we report the average of the ten consecutive executed frames.


\paragraph{Setting up a benchmark scene}

To ensure consistency between setups when recording performance with different settings, a default benchmarking scene was hard-coded with one light source, one plane, and five spheres, as shown in Figure~\ref{fig:adding-skybox}.
Furthermore, we evaluated the ray tracer with three different canvas sizes: a) 720p (1280 x 720), b) 1080p (1920 x 1080), and c) 4K (3840 x 2160).


\subsection{Overall real-time performance}
\label{section:real-time-p}

To assess performance we compute the average number of frames rendered each second (FPS) during runtime.
This is achieved by recording timestamps using \textit{System.nanoTime()} at the start and end of each iteration of the render loop to calculate the elapsed time for each frame being synthesized.

\begin{figure}[t!]
\centering
  \includegraphics[width=\linewidth]{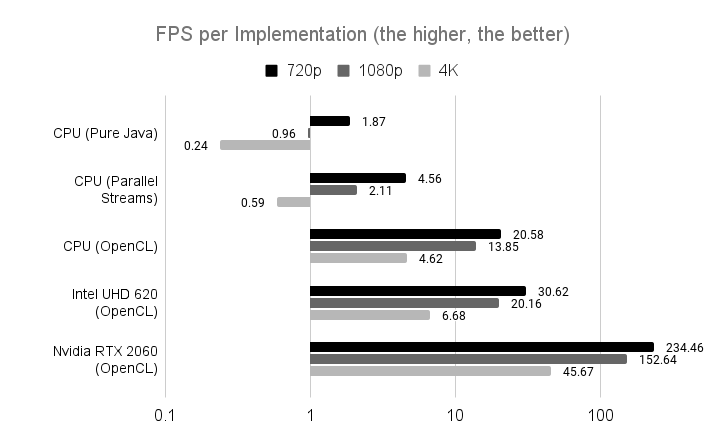}
  \caption{FPS achieved on different hardware.}
  \label{fig:performance-comp}
\end{figure}

Figure \ref{fig:performance-comp} depicts the average FPS of running the application with a shadow sample size of 1 (hard shadows) as well as a reflection bounce limit of 1.
These are the parameters of a fully classical Whitted-style ray tracer.

It can immediately be observed that the fully sequential execution of the application produces frame rates lower than 2 FPS at 720p and below 1 FPS for higher resolutions, which are unsuitable for real-time use cases (typically over 30 FPS). 
The discrete NVIDIA RTX 2060 GPU, however, has dominated this performance test with a 100-200x performance increase compared to sequential runs, achieving 2-4x higher frame rates at the 60 FPS mark.
Accelerating the application on the CPU via OpenCL, outperforms the sequential runs by 10-20x, while being 50\% slower than the Integrated GPU. 

\subsection{Obtaining ideal rendering parameters}

To achieve eye-pleasing visuals of the ray tracer, without compromising performance, the following rendering parameters have been examined:
\paragraph{Soft shadow sampling}
The most performance intensive setting proved to be the only parameter where our implementation diverged from classical Whitted-style ray tracing and adapted a distributed method for rendering soft shadows.
Unsurprisingly, as every single sampled shadow ray has to be checked with an intersection with every object in the scene, the number of expensive computations rapidly grow as we increase the sample size (Figure \ref{fig:shadow-sample-size}, the numbers in this chart are obtained with a reflection bounce limit of one).


As a middle-ground however, a shadow sample size of 200 proved to be enough to produce shadows without noticeable jags, while still achieving over 60 FPS.

\subsubsection{Reflection bounce limit}

\begin{figure}[ht] 
  \begin{subfigure}[b]{0.5\linewidth}
    \includegraphics[width=0.95\linewidth]{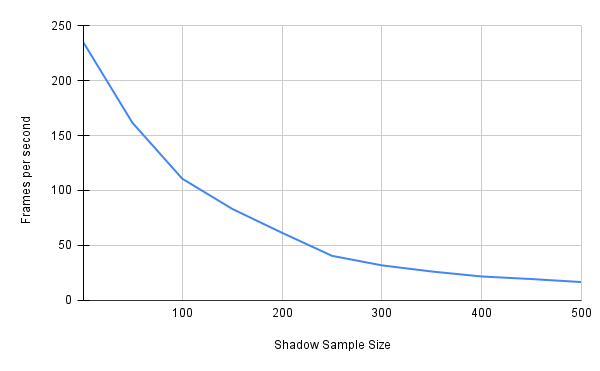}
    \caption{Increase in shadow sample sizes.}
    \label{fig:shadow-sample-size}
  \end{subfigure}
  \begin{subfigure}[b]{0.5\linewidth}
    \centering
    \includegraphics[width=0.95\linewidth]{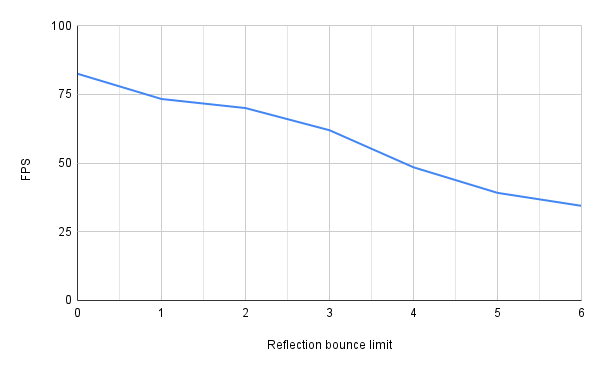}
    \caption{Increase in reflection bounce limits.}
    \label{fig:refl-limit-p}
  \end{subfigure}
  \caption{FPS achieved by increasing shadow sample sizes or reflection bounce limits.}\label{fps_fig}
\end{figure}

The analysis of reflection bounce limit was expectedly found to be highly dependent on the shadow sample size as the shadows within reflections are computed in the same manner as any initial view ray.
This means that increasing the reflection bounce limit, spawns more and more shadow rays, leading to more rapid performance decrease when the shadow sample size is high.
Looking at Figure \ref{fig:refl-limit-p}, we can observe that with a shadow sample size of 200, the performance steadily declines as the number of reflection bounces grows.
Visually however, reflection bounces larger than three were found to contribute an unnoticeable amount to the final visual, while resulting in the average framerate dipping under 60 FPS.


\subsection{Speedup of the rendering process}
\label{section:non-gui-p}

The full graphical application contains a number of overheads, outside of the rendering pipeline, that contribute to the final recorded performance, such as the efficiency of the \texttt{pixelWriter}, and the overheads of JavaFX. 
Thus, as a final evaluation, a non-GUI benchmark is set up to isolate and record the execution time of the \textit{Renderer::render} method by calling the function in isolation. 
This allows a speedup analysis of the rendering pipeline by itself showcasing the achieved performance improvements through the use of TornadoVM.

The non-GUI benchmark uses the same scene as the GUI evaluation, however the parameters for the shadow sample size and reflection bounce limit are set to 200 and 3 respectively as the ideal selected values.
The benchmark simply places the render function inside a loop of 100 iterations to generate 100 frames. 
The execution time of the loop is then divided by 100 to obtain the time taken to render 1 frame.
By doing this exercise, we observe speedups that range from 19.5x (720p - CPU/Integrated GPU) to 796x (4K - Dedicated GPU) compared to the Java Parallel Stream implementation.



  
\section{Related Work}
\label{section:relatedwork}

Since the first introduction of ray tracing~\cite{10.1145/1468075.1468082}, several implementations of different levels of complexity~\cite{whitted:1980, fuji} have been proposed throughout the years that utilize various data structures for improving the algorithmic performance~\cite{fuji, phdthesis, DBLP:journals/cgf/HapalaH11, 10.1145/360349.360354, 10.1145/965105.807481, article1, 10.1145/2492045.2492055}.
In addition, several studies that both assessed the performance of accelerated ray tracing algorithms\cite{10.1145/3023368.3023382} and proposed optimizations have been done over the years~\cite{10.1145/1572769.1572792, 10.1145/2492045.2492055, https://doi.org/10.1111/cgf.14410}.
Regarding commercial implementations, both NVIDIA and AMD offer ray tracing capabilities with dedicated ray tracing (RT) cores for acceleration~\cite{nvidia-pg:2022, amdgpuopen}.
Additionally, depending on the underlying algorithmic implementation of ray tracing, several hardware extensions have been proposed to accelerate various parts of the compute pipeline~\cite{10.1145/3104067}.

In programming terms, the majority of commercial ray tracing support is developed using heterogeneous frameworks such as CUDA~\cite{cuda:2022}, OpenGL~\cite{opengl:2022}, OpenCL~\cite{opencl:2022}, and Vulkan~\cite{vulkan:2022}.
Existing kernels or custom-built ones can be integrated with various programming languages via native bindings (e.g. Java Native Interface (JNI)).
In the context of Java, as in other managed programming languages, existing libraries can be employed via performing native calls and manually performing all memory allocation and copying.
To the best of our knowledge, this is the first work that adds high-performance graphics performance and ray tracing in Java, without having to utilize pre-built binaries or low-level external libraries.
\section{Conclusions \& Future Work}
\label{section:conclusions}

In this paper, we presented the first real time ray tracing framework written entirely in Java.
To achieve real time performance, the ray tracer has been implemented with TornadoVM that enables transparent hardware acceleration of Java programs on GPUs.
After analyzing the design of the Java ray tracer, we presented a comprehensive performance evaluation across different hardware architectures and frame resolutions.
Our results showcase that the Java-based ray tracer achieves real-time performance ranging between 45 and 234 FPS depending on the hardware accelerator and frame size.

Having established a baseline implementation as a proof-of-concept, the algorithmic details and complexity of the ray tracer can be augmented by integrated advanced features such as: reflection diffusing over different materials, refractions, anti-aliasing, polygonal meshes, and spatial partitioning.
In addition, the TornadoVM compiler can be enhanced to support the dedicated ray tracing cores currently present in modern GPUs via intrinsics or specialized instructions.

\begin{acks}
    This work is partially funded by grants from Intel Corporation and the European Union Horizon 2020 ELEGANT 957286. 
    Additionally, this work is supported by the Horizon Europe AERO, INCODE, ENCRYPT and TANGO projects which are funded by UKRI grant numbers 10048318, 10048316, 10039809 and 10039107.
\end{acks}

\balance
\bibliographystyle{ACM-Reference-Format}
\bibliography{references}

\end{document}